# Ultra-wideband statistical propagation channel model for implant sensors in the human chest

*A. Khaleghi*[1,2] *R. Chávez-Santiago*[1,3,*] *I. Balasingham*[1,3,*]

[1]*The Interventional Centre, Oslo University Hospital, Oslo, Norway*
[2]*Department of Electrical and Computer Engineering, K. N. Toosi University of Technology, Tehran, Iran*
[3]*Department of Electronics and Telecommunications, Norwegian University of Science and Technology (NTNU), Trondheim, Norway*
[*]*Institute of Clinical Medicine, University of Oslo, Oslo, Norway*
*E-mail: ali.khaleghi@rr-research.no*

**Abstract:** Implant medical wireless sensors for monitoring physiological parameters, automatic drug provision, and so on represent a new promising healthcare technology. Inherent characteristics of ultra-wideband (UWB) radio make this technology highly suitable for the wireless interface of implant sensors. A communication channel model is essential for developing these wireless systems. However, there are currently few models describing the radio propagation inside the human body. To address this problem, a statistical model is presented for UWB propagation channels inside the human chest in the 1–6 GHz frequency range. The proposed statistical model is developed from numerical simulations using a heterogeneous anatomical model that includes the frequency-dependent dielectric properties of different human tissues. Mathematical formulas for the computation of path loss, scattering and the statistical implementation of the channel impulse response at different depths inside the chest are described. Two typical depths for implanted sensors in the chest, namely 20 and 80 mm are analysed in detail. Average path loss of approximately 20 and 50 dB is observed in each case, respectively. Moreover, the channel exhibits little time dispersion with a root-mean-square delay spread below 1 ns in both cases. These results aim at facilitating the tasks associated with the design of in-body medical communication systems.

## 1 Introduction

In modern telemedicine systems, the physiological data of patients can be measured with the aid of electronic sensors located on and inside the human body [1]. The collected medical data are then transmitted wirelessly to an external unit for processing, thereby enhancing the health monitoring, diagnosis and therapy of the patients. Ultra-wideband (UWB) technology has great potential for applications in telemedicine, particularly for medical implant devices, owing to its inherent low power consumption, high transmission speed and simple electronics [2]. Although different frequency bands within the total 3.1–10.6 GHz allocated spectrum to UWB can be used in implant devices, the limitations due to high propagation loss through body tissues at higher frequencies must be taken into account. Our previous research work on UWB communications has demonstrated the feasibility of transmitting information at a high data rate from a capsule endoscope travelling inside the human digestive tract [3], and similar possibilities are expected in the chest.

Accurate knowledge of the propagation channel is necessary for an efficient design of UWB implant wireless communication systems. Considerable effort has been devoted to the characterisation of the propagation of UWB radio signals from on-body sensors; however, significantly fewer results are reported for the case of UWB implant devices. A number of path loss models for narrowband (NB) implant sensors are available in the literature [4–8] that regretfully are not suitable to model UWB propagation conditions. The IEEE 802.15.6 standardisation group [9] has issued several propagation models for medical and non-medical devices, both in-body and on-body, for wireless body area networks. Nevertheless, the in-body channel models therein characterise NB applications only, whereas the UWB models characterise on-body communication channels [10].

To the best of our knowledge, the sole UWB channel model for implant sensor communications that has been reported so far can be found in [11, 12]. In that work, the characterisation of an UWB link in the 3.4–4.8 GHz frequency band is presented. The model was developed assuming 20 arbitrary locations of a transmitting implant device inside the human chest between 6 and 18 mm depth.

This paper aims at providing a more general model, covering a broader frequency range and deeper locations inside the chest. For this sake, numerical simulations are conducted for evaluating the propagation channel for different depths inside the human body [13]. For each given depth several probes are uniformly distributed in a large square area inside the human chest. The pulse response of the in-body channels for the frequency range of 1–6 GHz was obtained discarding the antennas effects. A different approach to that presented in [11, 12] (and the other publications on UWB channel modelling [14]) is used herein. The channel impulse response is extracted using the so-called CLEAN algorithm [14–16]. Using the proposed channel model, the path loss and distortion of a transmitted pulse due to propagation through the human tissues in the chest area can be predicted accurately. Additionally, it can generate the power delay profile (PDP) and the delay spread characteristics with fair accuracy. These results are important for the design and performance evaluation of novel medical in-body wireless communication systems.

The rest of the paper is organised as follows: Section 2 discusses the in-body simulation scenario and the considerations used for the modelling of the wideband dielectric properties of the human tissues. Section 3 presents the characterisation of the path loss, scattering and channel impulse response. The implementation and validation of the proposed statistical model is described in Section 4. Finally, Section 5 summarises our conclusions.

## 2 Simulation scenario

The time-domain finite integration technique (FIT) was used to solve the Maxwell's equations for our numerical simulations. A model of the human body based on the visible human project of the National Library of Medicine (NLM®) [17] was incorporated in the simulations. Thereby a voxel representation of the human body with a resolution of 2 mm was embedded in the FIT simulator. The dielectric properties of the human tissues, permittivity and conductivity, were provided by Gabriel [18] based on the Cole–Cole model, which describes the frequency-dependent characteristics of the tissue materials. However, incorporating the highly complex material properties in the numerical simulations causes extremely long computational time [19]. Our own investigations [20, 21] revealed that a simplified equation for describing the frequency-dependent complex permittivity of the human tissues can be obtained with a second-order polynomial, which fits accurately the Gabriel's data by using

$$\varepsilon_r(\omega) = \varepsilon_r' - j\varepsilon_r'' = \varepsilon_\infty + \frac{\beta_0 + j\omega\beta_1}{\alpha_0 + j\omega\alpha_1 - \omega^2} \tag{1}$$

where $\varepsilon_r'$ and $\varepsilon_r''$ are defined as the real and the imaginary parts of the complex permittivity, $\varepsilon_r(\omega)$, respectively, $\omega$ is the angular frequency, and $\varepsilon_\infty$ is the permittivity at infinite frequency. $\alpha_0$, $\beta_0$, $\alpha_1$, $\beta_1$ are fitting parameters that are obtained by fitting to the Gabriel's four-pole Cole–Cole data using Newton method and least-square fitting. The fitting parameters were estimated for different tissue materials and the approximation in (1) was incorporated in the electromagnetic (EM) simulations.

The simulation scenario using the described anatomical model of an adult male is illustrated in Fig. 1. The chest is exposed to an incoming plane wave from the front. Ideal electric field probes are placed inside the chest within a rectangular prism space of 140 mm depth, 160 mm width and 80 mm height; the distance between contiguous probes is 20, 10 and 20 mm, in the aforementioned sides of the prism, respectively. There are 85 probes on each plane parallel to the skin (depth planes) for a total of 595 probes. The centre of gravity of the first depth plane is located at 5 mm inside and therefore the probes are embedded in different tissues.

The field probes are ideal frequency independent isotropic antennas with a specified polarisation. The probe arrays do not have any coupling among them. The electric field probes are considered to be in co-polar mode with the incident field polarisation, which is fixed along the $x$-axis (see Fig. 1). The incident plane wave is excited with the second derivative of a Gaussian pulse of 420 ps of time duration; this value was used in order to have most of the pulse energy within the 1–6 GHz frequency spectrum for a spectral bandwidth of –30 dB. The perfectly matched layer absorbing boundary condition is used for the simulations; thus the body environment reflections are disregarded.

## 3 Channel modelling

The received signal level at every probe location is obtained using numerical simulations as described above. The received pulses are distorted with respect to the transmitted ones due to the frequency-dependent loss properties of the body materials and the wave reflection from different tissues. Strictly speaking, the propagation channel consists of the propagation media, that is, the human chest tissues in this case, and the transmitter and receiver antennas. Different antenna structures lead to different impulse response of the propagation channels. Hence, for the sake of generality, the characterisation of only the propagation media is more convenient because this remains the same regardless of the types of antennas used for every specific design. Therefore the antenna effects are disregarded hereinafter. For specific applications, a link designer can compensate the effects of antennas by adding certain antenna-dependant link margins to the path loss calculations made with our model.

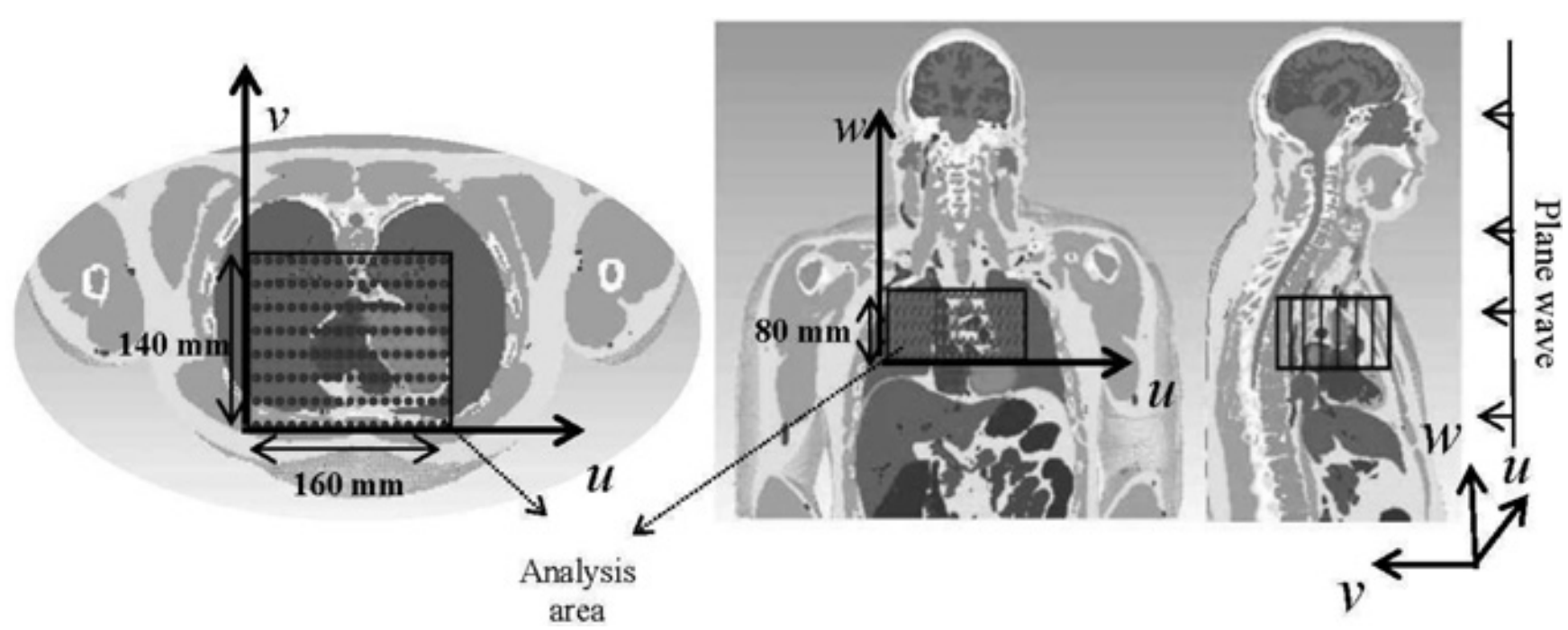

**Fig. 1** *Simulation scenario of the human torso with electrical field probes inside the chest; from left to right: top view, front view and side view, respectively*

In the following subsections the mathematical models for the path loss and scattering are presented; the statistical description of the pulse distortion for different depths inside the body is described too.

### 3.1 Path loss and scattering

The path loss is a measure of the average attenuation of the signal from transmitter to receiver. Owing to the plane wave excitation, the transmitted signal energy is uniform on the body surface. The total received energy at each probe position is obtained by integrating the Poynting vector during an observation time interval. This interval is sufficiently large to include 99% of the signal energy. Although the probes inside the chest are arranged in several planes that form a cubic shape as described previously, the contour of the chest surface is not a straight line. Therefore the path loss at each individual probe is computed and plotted against its actual depth inside the body from the skin (dots in Fig. 2). Then, a depth-dependent equation is fitted to the obtained data (solid line in Fig. 2). The path loss variations around the average value are referred to as scattering. These variations are caused by the different material dielectric properties along the propagation path. The statistical characterisation of the scattering is necessary for the calculation of the link margin. The probability density function (PDF) of the scattering (in decibels) in this case is well approximated by a Gaussian distributed random variable (RV), $\mathcal{N}$, with zero mean $\mu = 0$ and standard deviation $\sigma = 7.84$ (see Fig. 3). Hence, the mathematical expression of the best fitted line in dB-scale to the propagation loss (in decibels) as a function of the depth (in millimetres) including the scattering term is given as

$$L_{[db]}(d) = L_{0[db]} + a(d/d_0)^n + \mathcal{N}(0, \sigma) \quad (2)$$

where $d$ is the depth from the skin in millimetres ($1 < d < 120$), $a$ is a fitting constant with a value of 0.987, $d_0$ is the reference depth fixed at 1 mm, $L_0$ is the loss cross point with a value of 10 dB and $n$ is an exponent equal to 0.85.

### 3.2 Channel impulse response

The output of the FIT numerical simulations is a collection of distorted pulses affected by the propagation channel. In order to obtain the channel impulse response (CIR) for every realisation (probe), the transmitted pulse must be subtracted from the simulator output. The UWB signal arriving at the receiver is the sum of scaled and delayed replicas of the transmitted pulse. Thus, the impulse response, $h(\tau)$, of the channel can be expressed as

$$h(\tau) = \sum_{k=1}^{N} \alpha_k \delta(\tau - \tau_k) \quad (3)$$

where $\alpha_k$ and $\tau_k$ are the gain and time delay of the $k$th multipath component (MPC), respectively, and $N$ is the total number of MPCs. We used the so-called CLEAN algorithm [14–16] to extract the amplitude and delay of the MPCs for each channel realisation. The CLEAN algorithm first finds the largest pulse by correlating the received signal with a known template. The amplitude and time delay of the thus-identified pulse are recorded, then its power contribution is subtracted from the original signal at its corresponding time delay. The process is repeated until the power of the extracted amplitude compared to the power of the largest amplitude falls below a defined threshold, which in our case is fixed to 20 dB. Since the antenna effects were disregarded, the transmitted pulse was used as the signal template. If the antenna effects were considered, then the transmitted pulse distorted by the transmit antenna must be used. The total power of the MPCs in every obtained CIR is normalised to unity. The amplitude and delay of all the MPCs for each channel are extracted. For every depth plane, the statistics of the first MPC peak, the second MPC peak, and so on in each of the 85 CIRs are analysed. In order to verify the accuracy of the CLEAN algorithm, we reconstructed the distorted pulses (i.e. the output of the FIT simulator) by using the obtained amplitude and delay coefficients and the template waveform. Fig. 4*a* shows a sample of a distorted signal and the corresponding reconstructed pulse. As can be seen, the two signals are almost identical. The amplitude and delay coefficients of

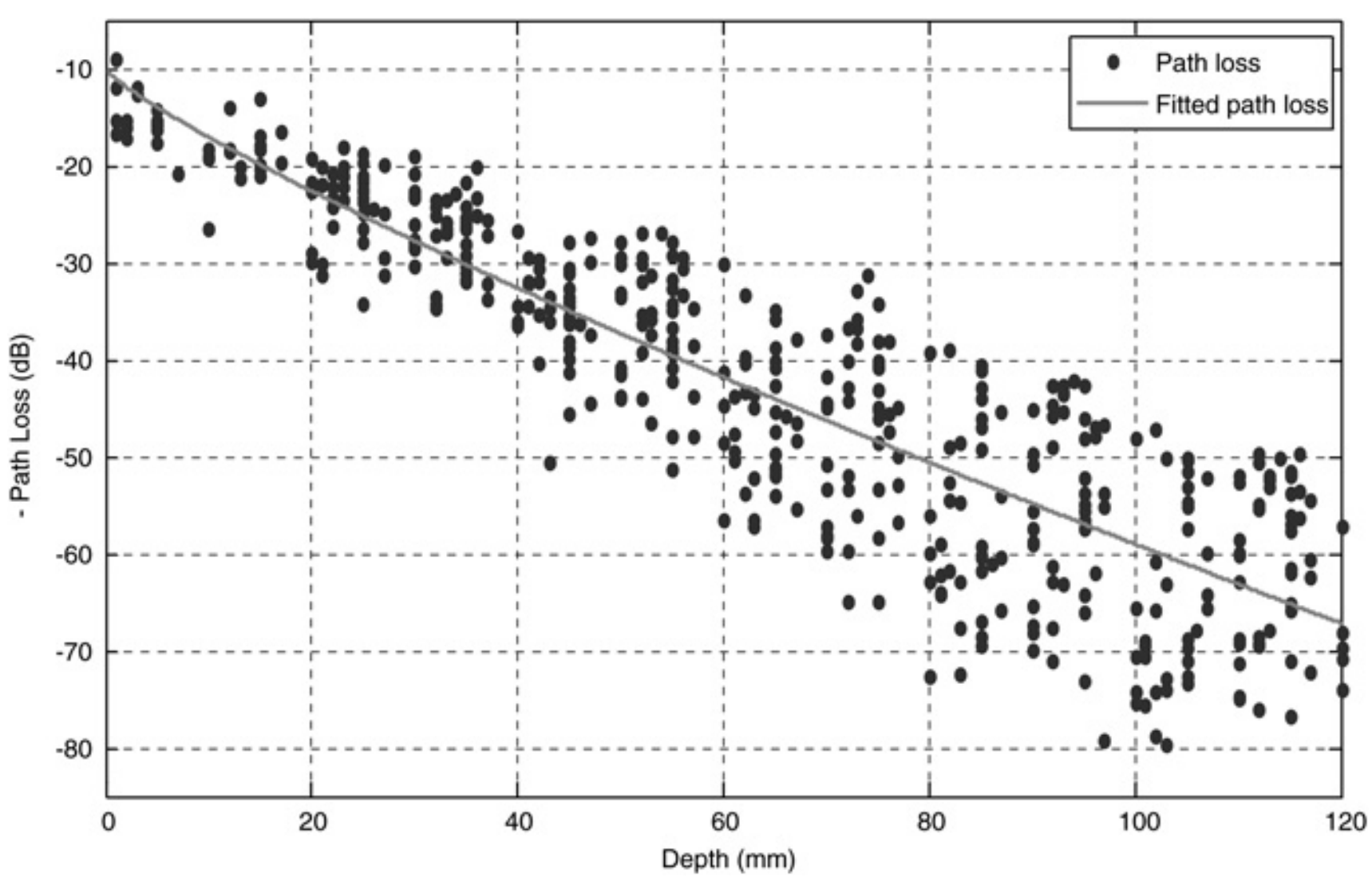

**Fig. 2** *Path loss (–dB) against depth (mm) inside the human chest*

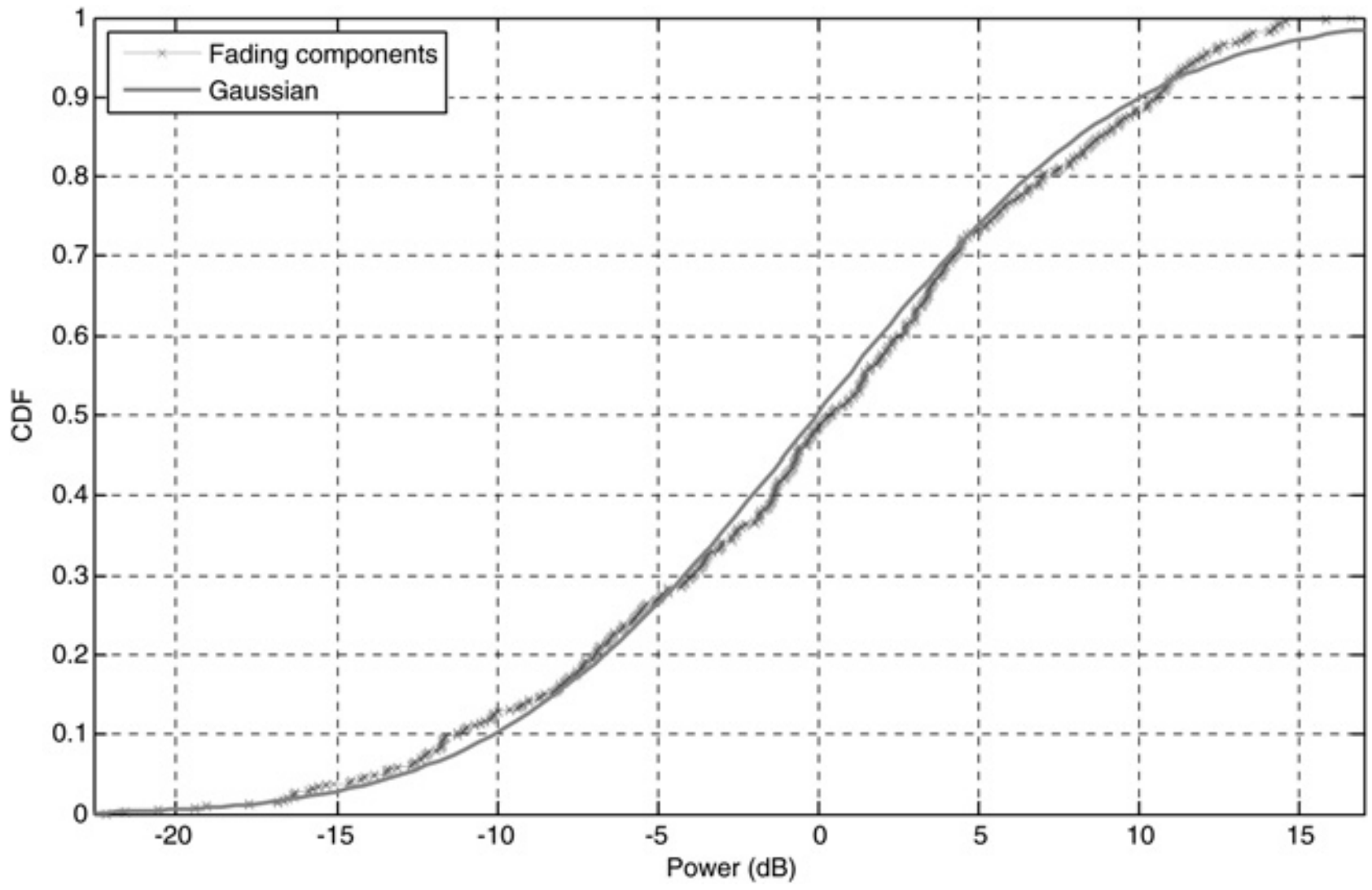

**Fig. 3** *CDF of the scattering inside the human chest; the fitted Gaussian curve confirms the log-normal distribution*

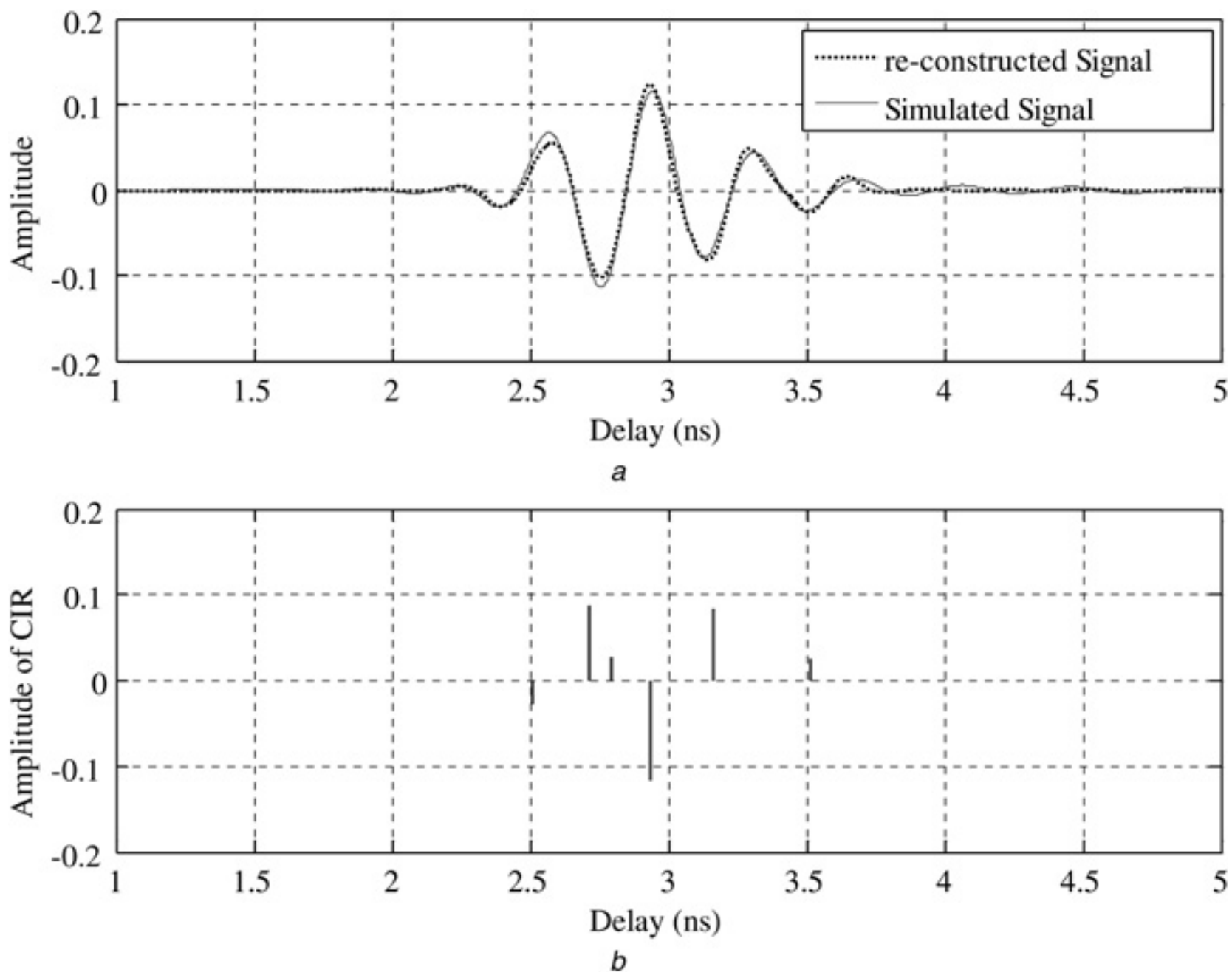

**Fig. 4** *CLEAN algorithm*

*a* Simulated signal at the output of the FIT simulator and the reconstructed signal by using the coefficients of the CLEAN algorithm
*b* Corresponding channel impulse response (amplitude and time delay) obtained with the CLEAN algorithm the corresponding MPCs are depicted in Fig. 4*b*. The correlation coefficient between the simulated signals and the re-constructed signals was larger than 0.95, which confirmed the accuracy of the CLEAN algorithm for the characterisation of the in-body channel. We note that the algorithm extracts the unlimited-band MPCs, thus even the non-resolvable MPCs for each channel realisation are obtained.

After obtaining the CIR, the next step in channel modelling is to provide the proper statistical description of the MPCs. A convenient characterisation of a multipath channel is to analyse the resolvable MPCs only [14, 22]. In this approach, the time scale is divided into small time intervals called bins. The bin size is generally chosen to be the resolution of the channel since two paths arriving within a bin cannot be resolved. The received power of non-resolvable MPCs is integrated within each bin to obtain the energy received as a function of the excess delay. The first bin corresponds to the first received MPC. In [22, 23] this approach is applied to characterise on-body channels. In contrast, we characterise statistically the unlimited band MPCs that were obtained with the CLEAN algorithm. The resolvable MPCs or bins are calculated later in order to compare the average power delay profile (APDP) and the root-mean-square (RMS) delay of the distorted pulses generated with the numerical simulations against those generated with the proposed statistical model. In our approach, the total power of MPCs for all the channel realisations are normalised to unity in order to remove the path loss and scattering effects. Then, the first, second and all the other relevant MPC peaks in the 85 channel realisations per depth are ‘aligned’ and the statistics of their power are investigated. After that, the statistics of the time delay are evaluated separately.

*3.2.1 Amplitude distribution of MPC peaks:* The distributions of the power of the MPC peaks were evaluated, showing that in the linear scale they follow a Gaussian distribution with parameters $\mu_{k,d}$ and $\sigma_{k,d}$ for the mean value and standard deviation of the $k$th MPC peak at a depth $d$. The Kolmogorov–Smirnov test was used for the confirmation of the Gaussian distribution for the MPC peaks power. Fig. 5 shows the cumulative distribution function (CDF) of the peak powers for the first, second and third MPCs. The CDFs of the corresponding fitted Gaussian distributions are also illustrated for comparison.

The correlation coefficients between the first and the rest of the MPC peaks were computed, which denoted low correlation (less than 0.3) between the first and the second peak. The correlation coefficient is even lower for the other peaks. This feature allows evaluating the statistics of the peaks as independent random variables (RVs). As $k$ increases the peak power diminishes and consequently the CDF curves are shifted to the left in Fig. 5. We can describe $\mu_{k,d}$ as a function of $k$ by an exponential equation as

$$\mu_{k,d} = \Omega_0(d)e^{-(k-1)\Lambda(d)} \tag{4}$$

where $\Omega_0(d)$ is the average power of the first peak and $\Lambda(d)$ is the decay exponent of the peaks for a given depth $d$. The values of $\Omega_0(d)$ and $\Lambda(d)$ for two typical depths, namely close to the skin ($d = 20$ mm) and deep inside the body ($d = 80$ mm), are given in Table 1.

Notice in Fig. 5 that the standard deviation of the first peak is significantly different to the standard deviation of the other MPCs, which are almost similar to each other. Therefore we provide separate standard deviation values for the first peak and the rest of the MPCs, respectively. The standard deviations of the other peaks can be considered as a constant value. This is summarised in Table 1 too.

*3.2.2 Time delay distribution of MPC peaks:* The time delay statistics for the first, the second and the other MPC peaks were evaluated. We observed that the distribution of the time delay for all the MPC peaks is Gaussian with parameters $\mu'_{k,d}$ and $\sigma'_{k,d}$ for the mean value and standard deviation of the $k$th MPC peak at a depth $d$. Fig. 6 shows sample CDFs of the time delay for the first and the second MPCs, respectively, at $d = 20$ mm. The CDFs of the corresponding fitted Gaussian distributions are also illustrated. From these results, we conclude that the average value $\mu'_{k,d}$ for all the MPCs can be considered as a constant for each given depth (see Table 1). However, the standard deviation can be expressed as an exponential function given by

$$\sigma'_{d,k} = \omega_0(d)e^{-(k-1)\lambda(d)} \tag{5}$$

The parameters of the exponential fitting function, $\omega_0(d)$ and $\lambda_0(d)$, are provided in Table 1 for $d = 20$ and 80 mm, respectively.

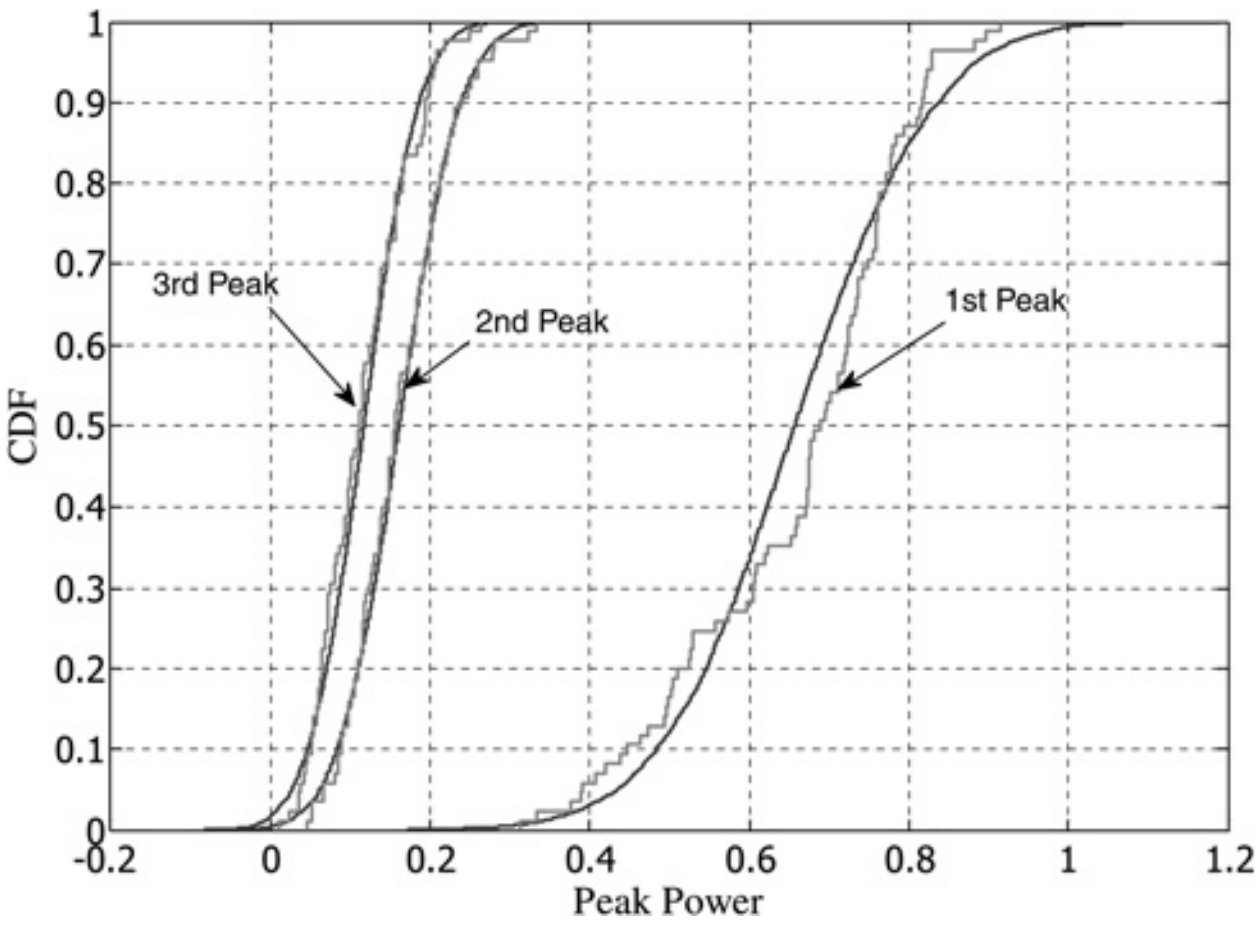

**Fig. 5** *CDF of the first, the second and the third peak powers and the corresponding Gaussian fitted CDFs for d = 20 mm*

**Table 1** Parameters of the statistical description of in-body channel for two different depths

| | Power | | Delay | |
|---|---|---|---|---|
| | $\mu_{k,d}$ | $\sigma_{k,d}$ | $\mu'_{k,d}$ | $\sigma'_{k,d}$ |
| $d = 20$ mm | $\Omega_0 = 0.65$, $\Lambda = 1.2$ | $\sigma_k = 0.14$ $k = 1$; $\sigma_k = 0.04$ $k > 1$ | 1.13 | $\omega_0 = 0.26$, $\lambda = 0.13$ |
| $d = 80$ mm | $\Omega_0 = 0.35$, $\Lambda = 0.5$ | $\sigma_k = 0.12$ $k = 1$; $\sigma_k = 0.02$ $k > 1$ | 2.29 | $\omega_0 = 0.37$, $\lambda = 0.09$ |

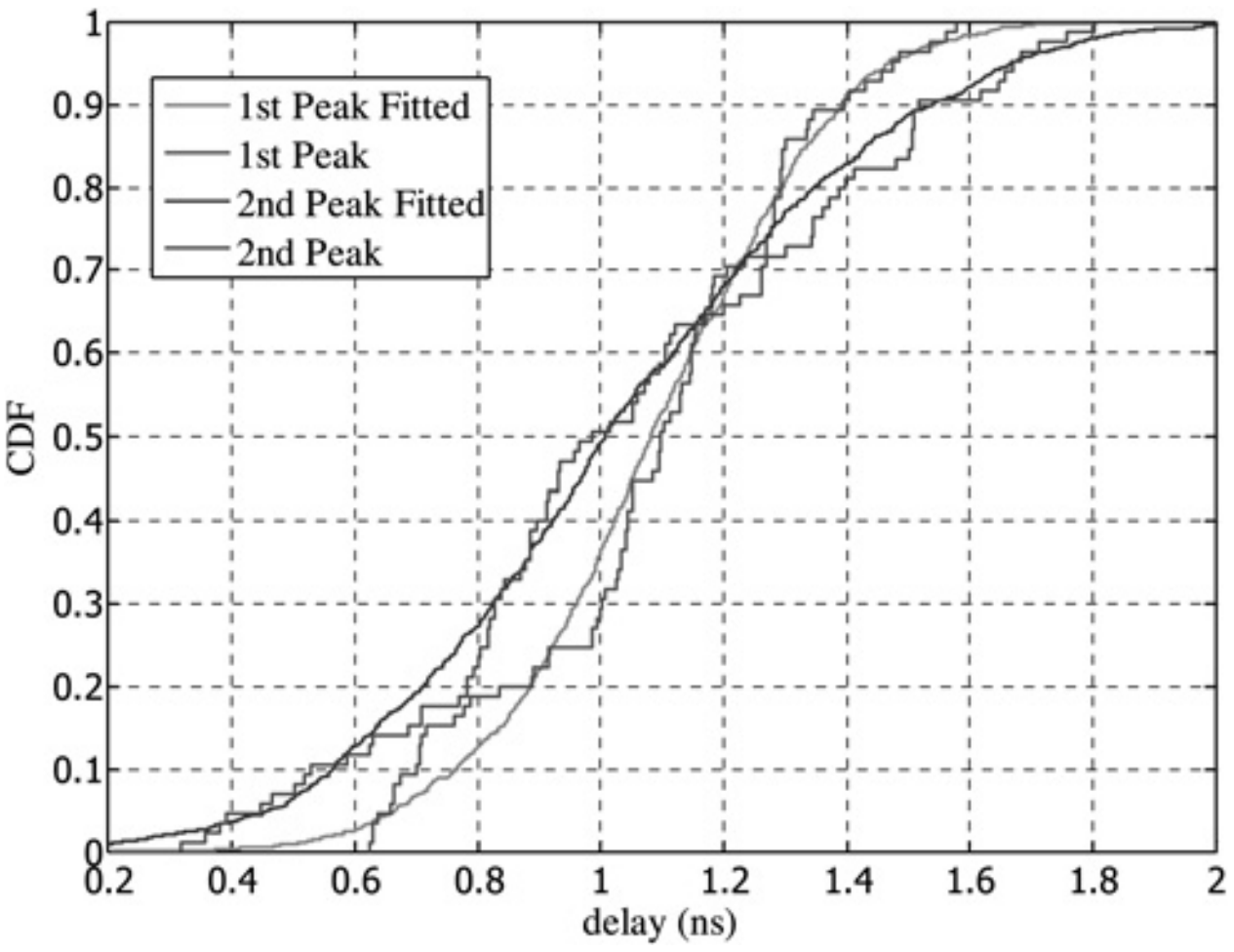

**Fig. 6** *CDFs of the time delay for the first and the second MPC peaks and the corresponding fitted Gaussian CDFs for d = 20 mm*

## 4 Model implementation

The statistical model of the CIR of UWB propagation through the human chest presented in the previous section can be easily implemented to reproduce distorted pulses, thereby bypassing the numerical simulations. For this sake, $N$ independent Gaussian RVs with parameters $(\mu_{k,d}, \sigma_{k,d})$ for the power $p_k = a_k^2$ and $N$ independent Gaussian RVs with $(\mu'_{k,d}, \sigma'_{k,d})$ for the time delay $\tau_k$ of the MPCs need to be generated. $N$ is chosen to be the number of MPCs containing significant energy. Through an extensive simulation campaign, we concluded that $N = 3$ for $d = 20$ mm and $N = 5$ for $d = 80$ mm are sufficient for accurate representation of the channel. The amplitude of the MPCs is the square root of the generated power values, $p_k$, which is multiplied by a factor $\beta_k$ that takes the values $+1$ or $-1$ with equal probability, that is

$$\alpha_k = \beta_k \sqrt{p_k} \tag{6}$$

Since the associated delay for each amplitude component is also generated, the construction of the CIR as described by (3) is straightforward.

Two comparison criteria are commonly used to verify the accuracy of channel models, namely the APDP and the RMS delay spread. In order to obtain the PDP of each channel realisation, we divide the time scale in bins of 420 ps under the assumption that two MPCs within a bin cannot be resolved. This resolution value (duration of a bin) is the reciprocal of the pulse bandwidth multiplied by the Hamming window factor, which was used in the numerical simulations. Then the APDP is calculated by averaging the PDP of all the channel realisations. The APDP of the simulated channels with the FIT tool are illustrated in Figs. 7*a* and *b* for $d = 20$ and 80 mm, respectively. About 85 channel realisations using the proposed statistical model were generated for the two considered depths, respectively, and their corresponding APDPs were computed. The comparison of the APDPs obtained with the FIT simulations and with our statistical model is shown in Fig. 7. As can be seen, in both cases the APDP obtained with the statistical model is very close to that obtained through time-consuming numerical simulations.

The RMS delay spread, which is the square root of the second central moment of the PDP, was computed for each channel realisation generated with both numerical and statistical simulations. The comparisons of CDFs of the RMS delay spread for the two different depths are shown in Figs. 8*a* and *b*, respectively. Once again, the statistical model approximates fairly well the numerical simulations. These tests demonstrate a good level of confidence in the model accuracy. Additionally, notice that in this case the propagation channel exhibits little RMS delay spread of around 1 ns in both cases. This contrasts with other propagation environments in which UWB communication links can be used. For instance, typical indoor propagation channels exhibit RMS delays in the order of 30–80 ns. It is obvious that the electronic design of the communication system is different for these two cases. An example of the special design for UWB in-body communications can be found in [3], where a single-path correlation receiver is proposed.

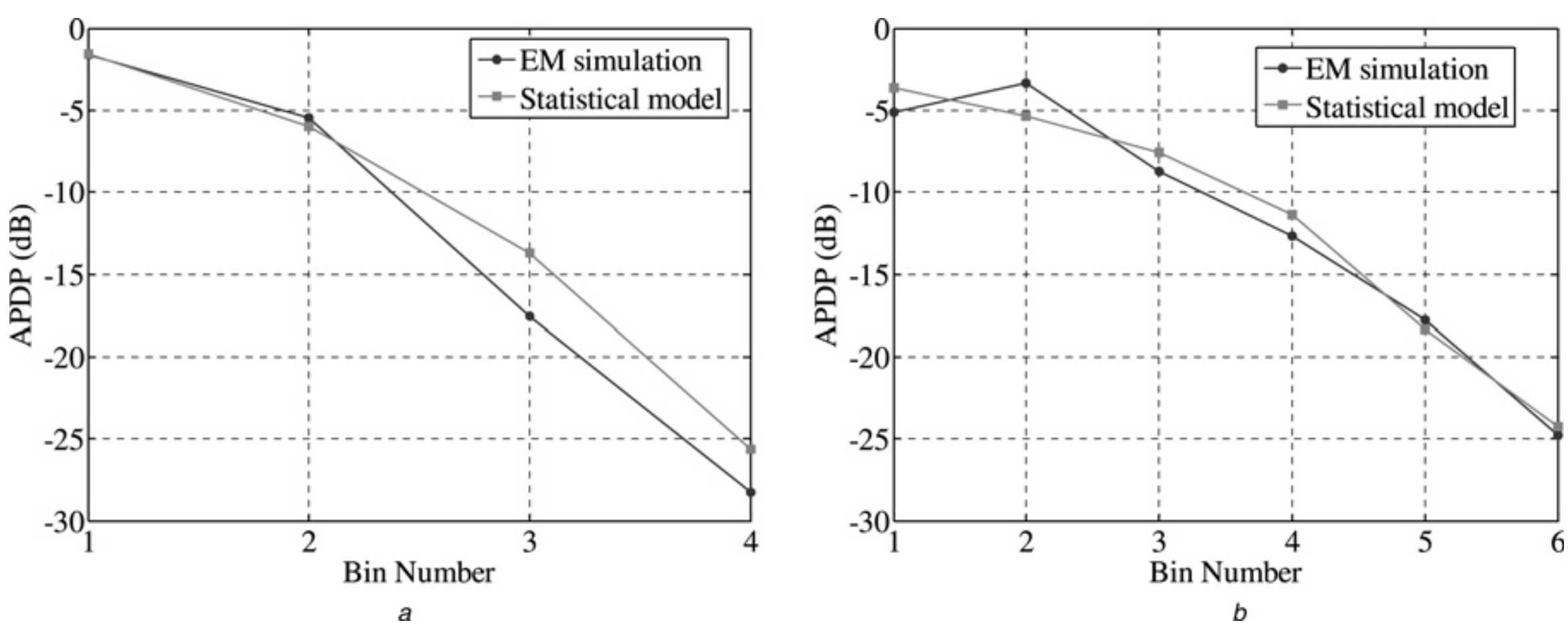

**Fig. 7** *Comparison of APDP derived from the EM simulations and the proposed statistical model*

*a* $d = 20$ mm
*b* $d = 80$ mm

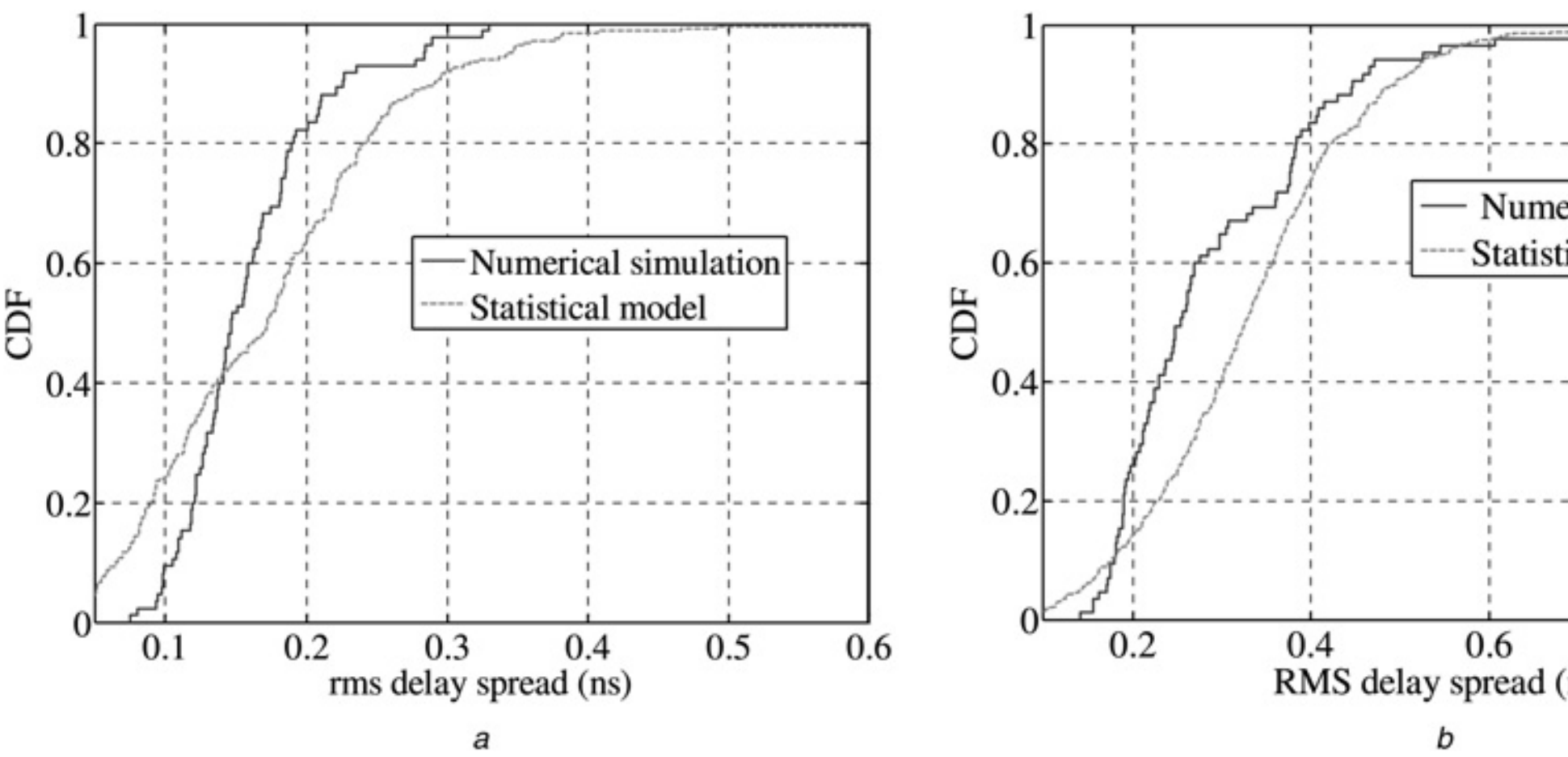

**Fig. 8** *Comparison of the RMS delay spread derived from the EM simulations and the proposed statistical model*

*a* $d = 20$ mm
*b* $d = 80$ mm

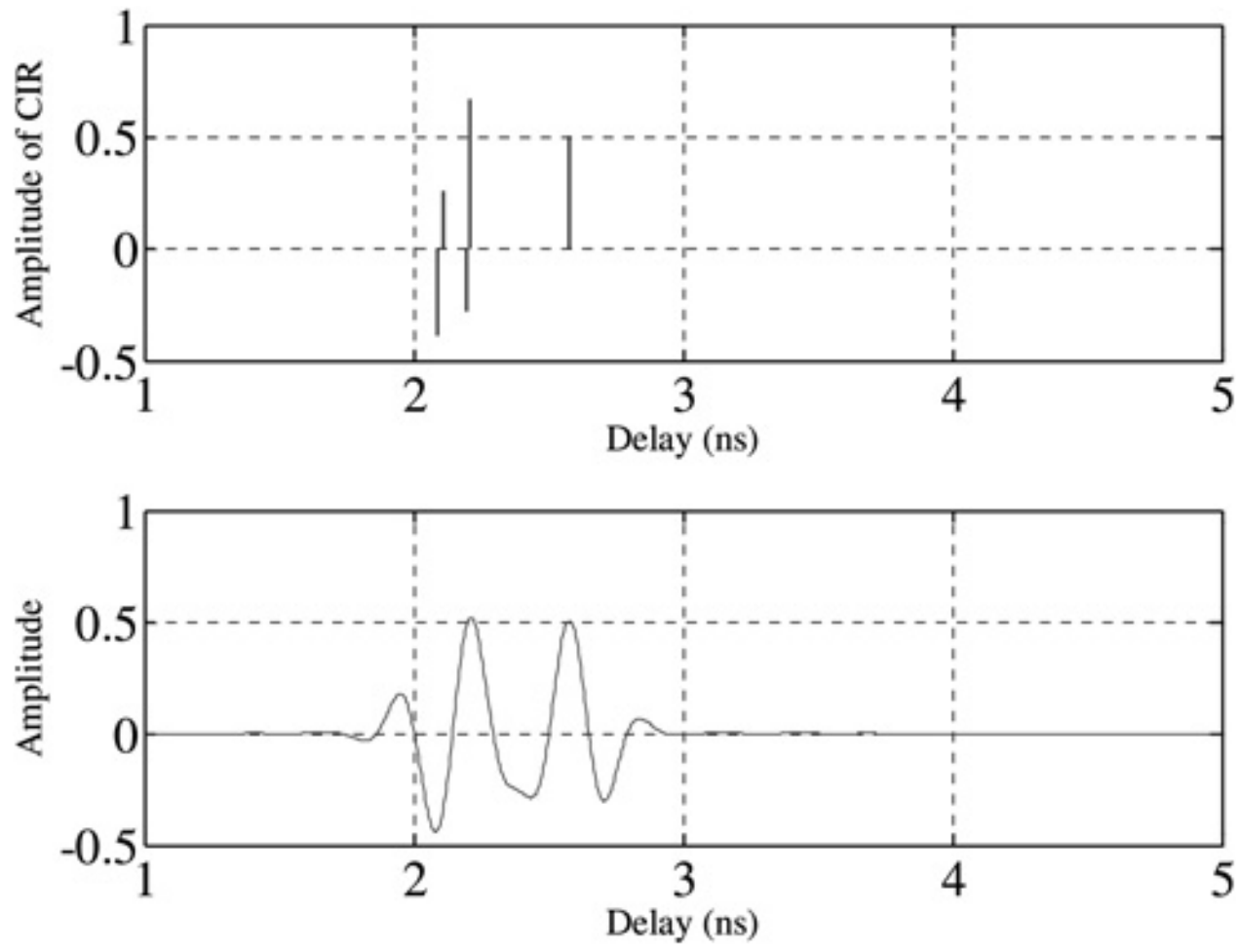

**Fig. 9** *Sample channel impulse response (top) and its corresponding distorted pulse shape (bottom) for d = 80 mm inside the chest*

Once the statistical channel model has been implemented, the resulting pulses after the distortion caused by the wave propagation through human tissues can be reproduced. The distorted pulses shape is essential for the design and evaluation of UWB transceiver systems, and the statistical model facilitates complicated tasks, such as template design for UWB receivers, bit-error rate estimation, design of synchronisation schemes, and so on. In order to generate the distorted pulse shapes, the coefficients of (3) are calculated according to the guidelines described previously. Then, the template signal (transmitted pulse) is convolved with the CIR and the distorted pulse shape is obtained. The total power of the distorted pulse can be scaled by using (2). Fig. 9 shows a sample of a CIR for $d = 80$ mm generated with the proposed statistical model and the distorted pulse shape it produces.

## 5 Conclusion

We have proposed an UWB time-domain channel model of the in-body propagation links through human tissues in the chest region for the frequency range of 1–6 GHz. This model is based on the statistical characterisation of multiple channel realisations obtained with time-consuming numerical simulations, which include the frequency-dependent material properties of the human tissues. For this sake, a complex heterogeneous anatomical model was embedded in the numerical simulation tool. Based on the data derived from the numerical simulations, the path loss as a function of depth inside the chest was modelled by a fitted mathematical function. The scattering term, which accounts for the path loss variations caused by different material properties of the tissues for a given depth, is included as a random variable. The channel impulse response was obtained with the so-called CLEAN algorithm. We demonstrated the efficiency of this algorithm for modelling in-body propagation channels. We validated the accuracy of the proposed statistical channel model by comparisons of the average power delay profile and the root-mean-square delay spread obtained with both the numerical simulations and the statistical model. The latter facilitates the reproduction of distorted pulses after propagating through human tissues, which is essential for the design and evaluation of UWB transceivers. Only the chest was considered in this paper and, owing to the highly inhomogeneous structure of the human body, a customised model must be derived for other anatomical parts. This work, however, has established some guidelines for the development of such models.

## 6 Acknowledgments

This work was supported by the MELODY Project, which is funded by the Research Council of Norway (contract no. 187857/S10). The authors are thankful to the anonymous reviewers for their insightful comments and suggestions to improve this paper.